\documentclass[fleqn,usenatbib]{mnras}

\usepackage{newtxtext,newtxmath}

\usepackage[T1]{fontenc}

\DeclareRobustCommand{\VAN}[3]{#2}
\let\VANthebibliography\thebibliography
\def\thebibliography{\DeclareRobustCommand{\VAN}[3]{##3}\VANthebibliography}

\usepackage{graphicx}	
\usepackage{amsmath}	
\usepackage{orcidlink}
\usepackage{tablefootnote}

\title[Optical polarization properties of AT 2023clx]{Optical polarization properties of the closest tidal disruption event AT 2023clx indicate origin from tidal stream shocks}

\author[K. I. I. Koljonen et al.]{Karri I. I. Koljonen$^{\orcidlink{0000-0002-9677-1533}}$,$^{1}$\thanks{E-mail: karri.koljonen@ntnu.no}
Kari Nilsson,$^{2}$
Ioannis Liodakis$^{\orcidlink{0000-0001-9200-4006}}$,$^{3,4,2}$
Elina Lindfors$^{\orcidlink{0000-0002-9155-6199}}$$^{2,5}$
\\
$^{1}$Institutt for Fysikk, Norwegian University of Science and Technology, H{\o}gskloreringen 5, Trondheim, 7491, Norway\\
$^{2}$Finnish Centre for Astronomy with ESO (FINCA), University of Turku, Finland \\
$^{3}$Institute of Astrophysics, Foundation for Research and Technology - Hellas, Heraklion, GR7110, Greece\\
$^{4}$NASA Marshall Space Flight Center, Huntsville, AL 35812, USA\\
$^{5}$Department of Physics and Astronomy, University of Turku, FI-20014 Turku, Finland\\
}

\date{Accepted XXX. Received YYY; in original form ZZZ}

\pubyear{\the\year{}}

\begin{document}
\label{firstpage}
\pagerange{\pageref{firstpage}--\pageref{lastpage}}
\maketitle

\begin{abstract}
Polarization observations of tidal disruption events offer unique insights into the accretion processes around supermassive black holes. Here, we present optical polarization observations of the nearby event AT 2023clx, obtained using the Nordic Optical Telescope. Our observations reveal a rise and subsequent decay in the polarization degree, temporally offset from the peak of the optical light curve, reaching maximum intrinsic polarization degree of $\sim$5 per cent. In addition, the polarization angle shifts by $\sim60^\circ-100^\circ$ between 6 to 20 days after the optical peak, remaining stable thereafter. Remarkably, the observed polarization variability closely resembles that of AT 2020mot, strongly suggesting a common mechanism for accretion disk formation in these events. The variability in both polarization degree and angle supports models in which tidal stream shocks drive the optical outburst during the accretion disk formation.
\end{abstract}




\section{Introduction}

Tidal disruption events (TDEs) occur when a star passes close enough to a supermassive black hole that the tidal forces exerted by the black hole exceed the star's self-gravity, leading to its destruction. A portion of the disrupted stellar debris becomes bound to the black hole’s gravitational potential, forming elongated streams of matter that collide and eventually settle into an accretion disk around the black hole \citep{rees88,evans89}. The mass accretion rates in TDEs can exceed the Eddington limit \citep{strubbe09,lodato11}, making TDEs bright sources in the local Universe. Thus, TDEs provide unique opportunities to study accretion disk formation and associated high-energy phenomena, such as relativistic jets, around supermassive black holes.

The increasing discovery rate of optically bright but X-ray-dim TDEs -- commonly referred to as \textit{optical TDEs} \citep{gezari12,vanvelzen21} -- has sparked debate about the mechanisms driving their observed emission. Many optical TDEs also exhibit late-time detections in the X-ray \citep[e.g.,][]{pasham17,guolo24} and radio \citep{horesh21,cendes24}, suggesting delayed accretion disk and jet formation or the obscuration of X-ray emission. Several models have been proposed to explain the optical flares in TDEs, including the reprocessing of X-ray emission from the accretion disk by stellar debris \citep[e.g.,][]{loeb97,strubbe09,guillochon14,roth16}, radiatively or collisionally driven winds \citep{metzger16,parkinson22,lu20}, and shock-powered emission from the inefficient circularization of disrupted stellar material \citep{shiokawa15,piran15,jiang16,ryu23,price24,steinberg24}. 
While these scenarios predict similar optical light curves (and late-time X-ray/radio behavior; \citealt{gezari17}), they are expected to differ in their polarization signatures---specifically in the polarization degree, and variability of the polarization angle. Reprocessing models predicts polarization degrees less than 14 per cent (at most preferable conditions) and minimal, slow variability for the polarization degree and angle \citep{leloudas22,charalampopoulos24}, while shocks can yield much higher polarization degree and exhibit strong variability, as seen in many astrophysical scenarios with ordered magnetic fields (e.g., in blazars; \citealt{angel80,hovatta19}, and gamma-ray bursts; \citealt{mundell13}). This makes polarization a novel diagnostic tool for probing emission mechanisms in TDEs.

Recent observations have revealed a wide range of polarization properties in optical TDEs \citep{leloudas22,patra22,liodakis23,koljonen24}. Notably, AT 2020mot exhibited a very high intrinsic polarization degree ($\sim$ 25 per cent; \citealt{liodakis23}) implying shock-powered emission, as reprocessing scenarios predict maximum polarization degrees of only 6-14 per cent \citep{leloudas22,charalampopoulos23}. Furthermore, fast variability in both polarization degree and angle has been observed during peak optical emission, with possible explanations including shocks or a precessing accretion flow \citep{koljonen24}.

\subsection{AT 2023clx}

AT 2023clx (ASASSN-23bd) is a nuclear transient event identified as a TDE in the nearby galaxy NGC 3799 \citep{taguchi23,johansson23}, which appears to be merging with its companion galaxy, NGC 3800 \citep{ramospadilla20}. NGC 3799 is a star-forming galaxy classified as a low-ionization nuclear emission-line region (LINER) galaxy \citep[see discussion in][]{charalampopoulos24}, indicative of either low-luminosity AGN activity \citep{ho08}, a population of post-AGB stars \citep{yan12}, or merger-induced shocks \citep{kewley15}. 

Previous studies have extensively examined the possibility of AGN activity as the origin of AT 2023clx \citep{charalampopoulos24,hoogendam24,sfaradi23} but have deemed it unlikely due to several factors: the estimated low mass of the black hole ($M_{\rm BH} \sim 10^6 M_{\odot}$), the event's fast rise and shallow decline in the optical, the absence of stochastic variability, no detectable pre-event variability in archival optical or X-ray data, the lack of Mg II or Fe II lines in spectra, the decreasing emission line widths as the event decays, flat Balmer decrements, the similarity of the optical spectra to other TDEs, and late-time X-ray detection \citep{charalampopoulos24,hoogendam24}. AT 2023clx has also shown potentially transient radio emission with 0.4 mJy detection $\sim$5 days post-peak, and a pre-event 3$\sigma$ upper limit of 0.37 mJy \citep{sfaradi23}. 

Located at $z=0.01107$ \citep{albareti17}, NGC 3799 hosts the closest TDE observed to date \citep{zhu23}. In addition, AT 2023clx presents unusually fast evolution and faint brightness, which may suggest the disruption of a low-mass star. This event hints at the existence of a broader population of faint TDEs currently undetectable with existing facilities \citep{charalampopoulos24}.

In this paper, we present optical polarization observations of AT 2023clx obtained with the Nordic Optical Telescope (NOT). In Section 2, we describe the observations and the polarization data analysis. In Section 3, we present the polarization results during the optical outburst decay phase of AT 2023clx. In Section 4, we discuss these results, particularly in the context of the tidal stream shock scenario, and compare them to the similar evolutionary behavior observed in AT 2020mot. Finally, we provide our conclusions in Section 5.

\section{Observations and data analysis}  

\begin{table}
\caption{\label{polobslog}Log of the NOT/ALFOSC polarization observations.}
\begin{center}
\begin{tabular}{llll}
(1) & (2) & (3) & (4)\\
\hline
  YYYY-MM-DD & UT$^a$ & Band & t$_{\rm exp}$(s)$^b$\\
\hline
2023-02-28 & 00:08:58 & B & 300\\
           & 00:56:34&  V & 200\\
           & 00:30:48 & R & 300\\
\ \\           
2023-03-14 & 04:17:28 & B & 300\\
           & 04.44:56 & V & 250\\
           & 03:54:11 & R & 300\\
\ \\
2023-03-23 & 01:53:20 & B & 300\\
           & 01:29:02 & R & 300\\
\ \\
2023-03-29 & 23:50:25 & B & 300\\
           & 23:12:16 & R & 300\\     
\ \\
2025-02-25 & 02:46:32 & B & 300\\
           & 06:20:31 & V & 300\\                
\hline
\end{tabular}\\
\end{center}
Notes :\\
$^a$Time of the start of the first exposure.\\
$^b$Exposure time per lambda plate position.
\end{table}

We observed AT 2023clx using the Alhambra Faint Object Spectrograph and Camera (ALFOSC) with the Filter and Shutter Unit Polarizer (FAPOL) over five epochs: four starting near the peak optical brightness and continuing for approximately a month during the optical decay, and a fifth taken two years after the event for calibration purposes. The observation log is provided in Table \ref{polobslog}. FAPOL uses a calcite crystal to split incoming light into two orthogonally polarized beams. A $\lambda/2$ plate, placed in the optical path before the calcite, rotates the orientation of the incoming light relative to the instrument. Each polarization observation consisted of four exposures taken at $\lambda/2$ plate angles of 0$^{\circ}$, 22.5$^{\circ}$, 45$^{\circ}$ and 67.5$^{\circ}$. Polarization parameters were calculated from the flux ratios of the two beams, hereafter referred to as the o- and e-beams, as follows: First, we measured the net counts from the o- and e-beams within a 3.0~arcsec radius aperture in each image, denoted as $N_i^o$ and $N_i^e$ for $i=0^{\circ},22.5^{\circ},45^{\circ},67.5^{\circ}$. We then
computed the flux ratios as $Q_i = N_i^o/N_i^e$. These ratios were used to calculate the normalized Stokes parameters
$q$ and $u$:
$$
q = \frac{Q_1 - Q_3}{Q_M}, \ u = \frac{Q_2 - Q_4}{Q_M},
$$
where $Q_M = Q_1 + Q_2 + Q_3 + Q_4$. From these, the
degree of polarization $p$ and polarization position angle $PA$ were derived as:
$$
p = \sqrt{q^2 + u^2}, \  \ PA = 0.5\times\arctan{\Big(\frac{u}{q}\Big)}.
$$
The position angle, PA, was calibrated using a previously determined instrumental offset derived from observations of high-polarization standard stars\footnote{The instrument’s stability is monitored regularly using high- and zero-polarization standard stars. The PA offset has been determined to be 91.6$^{\circ}$ and instrumental polarisation less than 0.3 per cent. We did not observe any deviations in instrument performance in the standard star observations taken close in time to our science observations.}. The degree of polarization $p$ was then de-biased using the equation:
$$
p_0 = \sqrt{p - 1.41 \times \sigma_p},
$$
where $\sigma_p$ = $(\sigma_q + \sigma_u)/2$, and $\sigma_q$ and
$\sigma_u$ were calculated via Gaussian error propagation from $\sigma_{N_i^{o,e}}$ \citep{simmons85}. 

Additional observations were made on January 12th, 2025, during which we obtained direct images of the host galaxy of AT 2023clx through B, V, and R filters. The purpose of these observations was to create a pure host galaxy template after the TDE has faded, enabling accurate host galaxy subtraction from the polarization images. The telescope, instrument, and filters were the same as those used for the polarization observations, with the only difference being the absence of the $\lambda$/2 plate and the calcite. The exposure time was 240 s per filter, and the image quality was better than in most of the polarization data.

\subsection{Host galaxy correction}

\begin{figure}
\begin{center}
\includegraphics[width=0.23\textwidth]{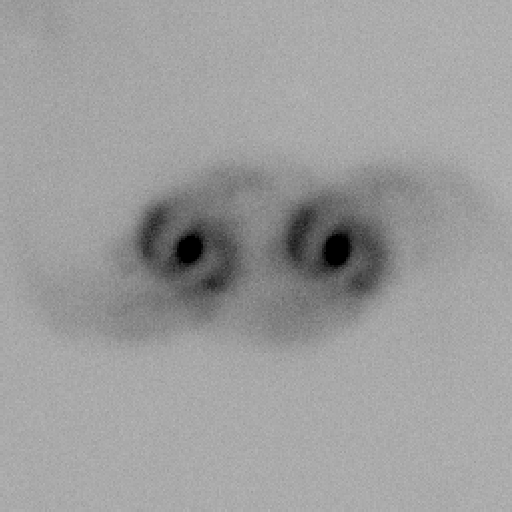}
\includegraphics[width=0.23\textwidth]{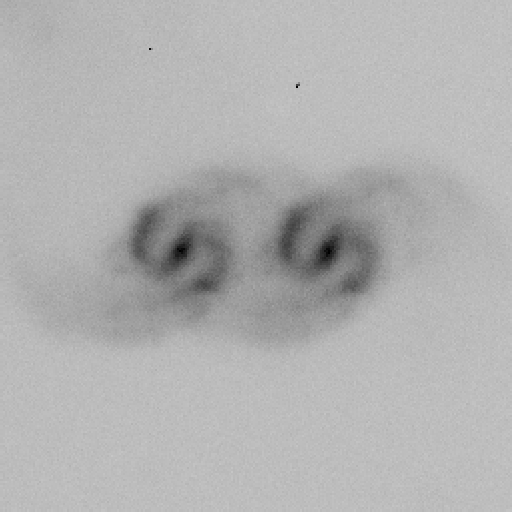}\\
\includegraphics[width=0.23\textwidth]{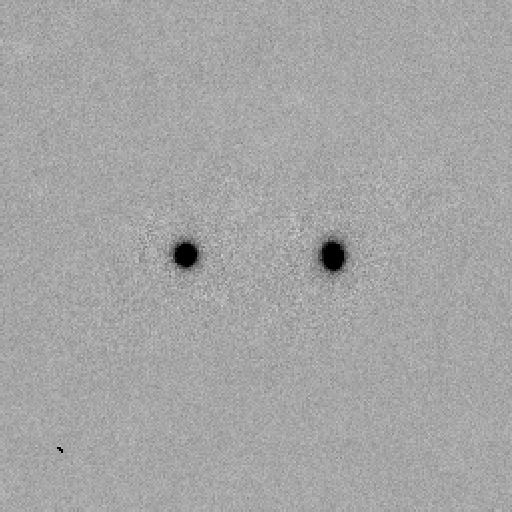}
\includegraphics[width=0.23\textwidth]{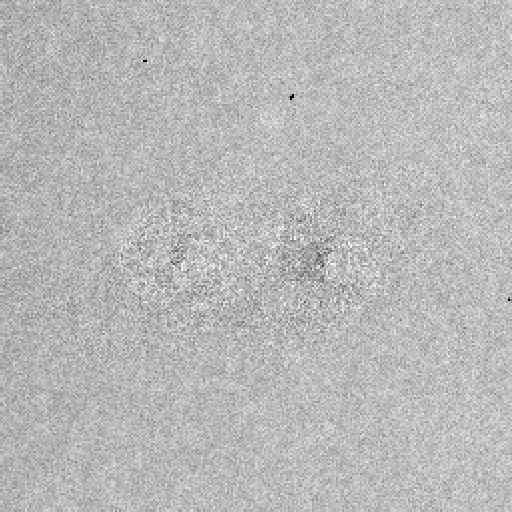}
\end{center}
\caption{Two examples of the host galaxy subtraction in the B-band. \textit{Left panels}: data obtained on 2023-03-23. \textit{Right panels}: data obtained on 2025-02-25, after the TDE had faded. \textit{Top row}: The observed image at a calcite position of 0$^{\circ}$. \textit{Bottom row}: Corresponding residual images after the host galaxy subtraction. North is up and East is to the left in all panels. The field of view is $52\times52$ arcsec.
\label{fig:polkuva}}
\end{figure}

Extracting the intrinsic polarization of the TDE requires subtracting the unpolarized contribution from the host galaxy within the source aperture. The host galaxy of AT 2023clx is a spiral galaxy with a surface brightness profile too complex for analytical modeling (see Fig. \ref{fig:polkuva}). Therefore, we used properly scaled template images obtained nearly two years after the TDE peak emission, as described in the previous section. These images were the used to construct a TDE-free polarization image in each band, which was fitted to the observed images. 

Several aspects of the polarization data complicate modeling, and careful procedures were necessary to achieve clean subtraction. First, the positions and separations of the o- and e-beams vary with filter. Second, the two beams have different point spread functions (PSFs) and transmit unequal amounts of light (the gain difference being one of the main reasons for taking four images rather than two). Third, the o- and e-beams are not perfectly aligned due to differences in their beam centers. To mitigate any leftover TDE contamination, we excluded the central 3.2~arcsec region around both beam centers. The outer boundary (11.8~arcsec) was selected to include the brightest regions of the host galaxy, while the inner exclusion ensured the TDE signal was fully removed.

Since the template images had better resolution (i.e., smaller full-width half maximum; FWHM) than the polarization images, we matched their resolutions by convolving the templates with elliptical Gaussian kernels. The convolution kernel parameters (FWHM, ellipticity, and position angle) were free parameters in the fit, and separate kernels were used for the o- and e-beams. In the few cases where the polarization images had higher resolution than the templates, we first convolved the polarization images with a Gaussian to match the template resolution before proceeding with the fit.

The model was defined by the following 14 parameters: 
(1) \& (2) the x- and y-position of the o-beam image, 
(3) the separation between the o- and e-beams (fixed),
(4) the calcite misalignment (fixed at $0.4^{\circ}$), 
(5) template size scaling (fixed at 1.0075),
(6) o-beam template scaling, 
(7-9) o-beam kernel FWHM, ellipticity and position angle,
(10) e-beam template scaling, 
(11-13) e-beam kernel FWHM, ellipticity, and position angle, 
and (14) the sky level. 
The fits typically converged within 2000--3000 iterations. Given the number of free parameters, multiple fits with different initial conditions were performed to ensure convergence to the global chi-squared minimum.

The residual images after host subtraction (see Fig. \ref{fig:polkuva} left) show a clear point-source excess at the galaxy center, corresponding to the TDE. We performed aperture photometry on the host-subtracted images using a 2.5-arcsec aperture radius and calculated the polarization properties of the TDE using standard methods \citep[e.g.][]{landi07}. This aperture radius was chosen as a compromise between two goals: staying close to the optimal value of 1--1.5~FWHM ($\sim$ 1--1.5~arcsec), and including most of the TDE flux despite PSF differences between the two beams. 
Within a reasonable range of aperture radii (2--4~arcsec), the polarization parameters were stable within the uncertainties derived from photon statistics and readout noise.

The fitted template scaling factors also allow us to estimate the background polarization, which includes both instrumental and any potential host contributions. We measured a small but consistent polarization of P$_{\rm bg}$ = (0.46 $\pm$ 0.07) per cent at a position angle of PA = (136 $\pm$ 5)$^{\circ}$, consistently across all bands. We consider the majority of this signal likely to result from instrumental polarization and the template-fitting procedure, for which the polarization degree is expected to be higher than in the case of a point-like source observed with a circular aperture, where instrumental polarization is typically below 0.3 per cent.\footnote{We measured P$_{\rm bg}$ = (0.2 $\pm$ 0.15) per cent for a zero-polarization standard star observed close in time to our science observations.} 

However, we note that \citet{uno25} measured a similar polarization degree 121 days after the peak, which could indicate a constant level of polarization following the event arising from, e.g., emission from a formed accretion disk. In the following, we estimate if the background polarization could be due to accretion disk. Following \citet{mummery24}, the optical g-band luminosity of the accretion disk during the plateau phase for a black hole mass of $M_{\rm BH} = 10^6\,M_\odot$ is $L_g \approx 10^{41}\,\text{erg\,s}^{-1}$, which corresponds to an apparent magnitude of approximately 19.6 at a distance of 47.8 Mpc. The host galaxy has a g-band magnitude of $\sim$14.2 \citep{charalampopoulos24}, resulting in a flux ratio of $F_{\rm host}/F_{\rm disk} = 10^{-2.5(14.2-19.6)} \sim 145$. Therefore, even assuming a maximum polarization degree of 11.7 per cent for the disk, its contribution would only amount to $\sim$0.08 per cent of the total observed polarization. We therefore consider it highly unlikely that the measured background polarization originates from the accretion disk.

\subsection{Line depolarization}

\begin{figure}
    \centering
    \includegraphics[width=1.0\linewidth]{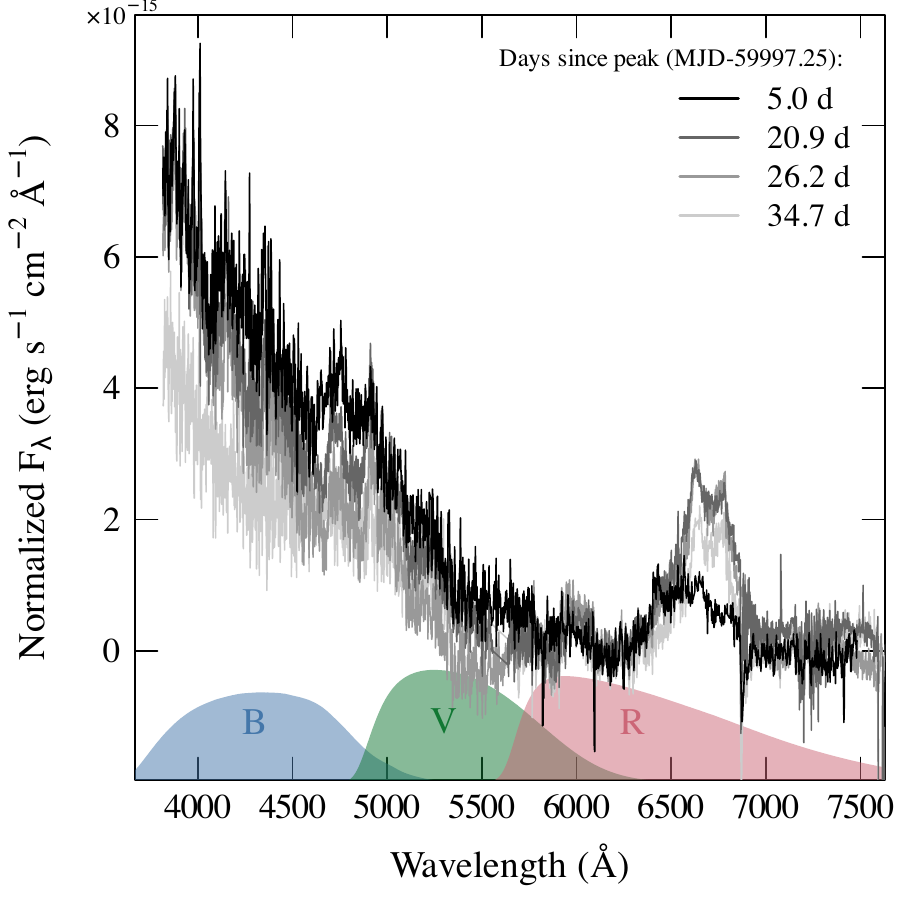}
    \caption{A collection of host-subtracted optical spectra from \citet{charalampopoulos24} close to our NOT polarization observations. The transmission curves of the NOT filters used are shown at the bottom. The H$\alpha$ line dominates the R-band flux which may affect the measured polarization degree significantly in that band.}
    \label{fig:specfig}
\end{figure}

AT 2023clx exhibits a broad H$\alpha$ line in its spectra \citep{charalampopoulos24}. If this line is unpolarized, the R-band may be affected by line depolarization (e.g., \citealt{leloudas22}; see Fig. \ref{fig:specfig}). Indeed, spectropolarimetric observations of AT 2023clx at 5.4 days post-peak show a dip in the polarization degree at the location of H$\alpha$, consistent with line depolarization \citep{uno25}. 

To estimate the potential impact of line depolarization in our filters, we perform a rough calculation by fitting optical spectra (taken from \citealt{charalampopoulos24}) obtained near the time of our polarization observations with an exponential continuum and three Gaussian profiles representing the strongest emission lines observed in the spectra: H$\alpha$, H$\beta$, and He II $\lambda$4686. Each component is then convolved with the corresponding filter transmission curve, and their fluxes within the filters are calculated.

The resulting line-to-continuum flux ratios are approximately 0.02–0.1 in the B-band and 0.3–1.2 in the R-band. The V-band is largely free from any strong emission lines. It is worth noting that the R-band ratio likely represents a lower limit, as the host galaxy’s flux increases at redder wavelengths, and it is challenging to accurately determine the level of the intrinsic TDE continuum in the spectra. Consequently, if the H$\alpha$ line is unpolarized, the intrinsic polarization degree in the R-band could be higher by up to a factor of two.

Therefore, we consider the R-band continuum polarization results to be potentially compromised during epochs when the H$\alpha$ line is strong (i.e., the last three epochs), as the exact depolarizing effect of the H$\alpha$ line remains uncertain. However, this primarily affects the polarization degree rather than the polarization angle. In contrast, the B- and V-band measurements are expected to provide a more accurate estimate of the polarization of the underlying continuum. 

\section{Results}

\begin{figure}
    \centering
    \includegraphics[width=1.0\linewidth]{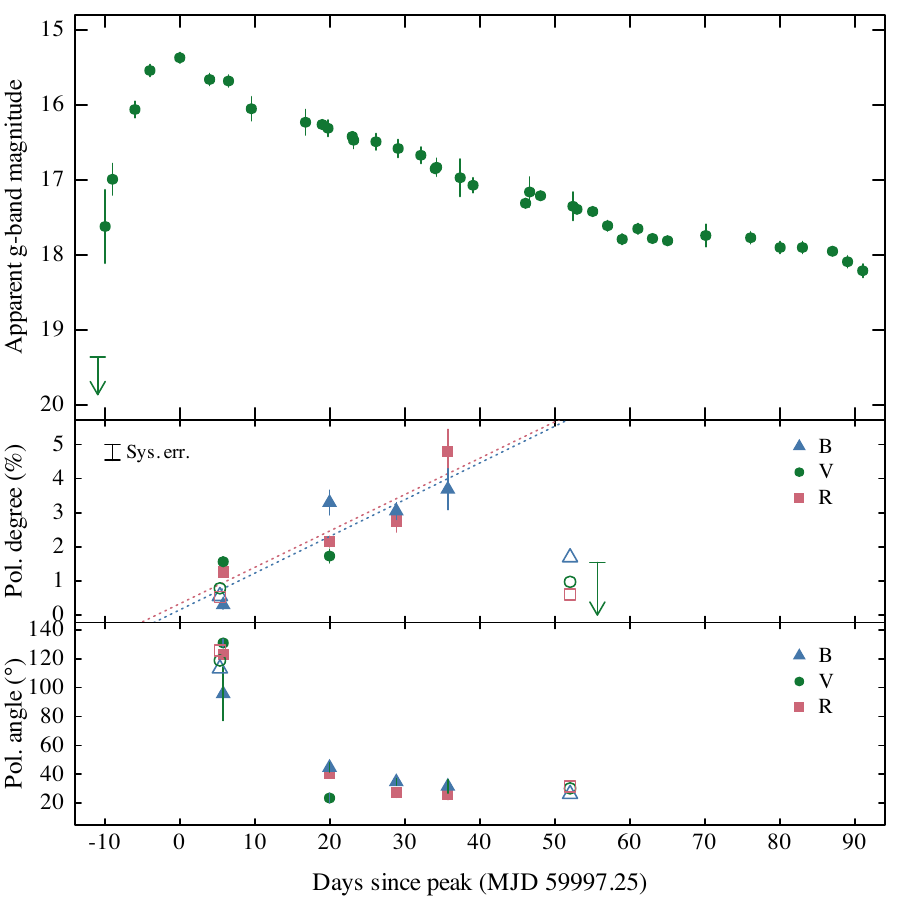}
    \caption{\textit{Upper panel:} ASAS-SN and ZTF g-band apparent magnitude light curve (data tabulated in \citealt{charalampopoulos24}). \textit{Middle panel:} Multi-band, host-corrected polarization degrees from our NOT observations (filled symbols), a 2$\sigma$ upper limit from the Liverpool Telescope (green arrow), and band-averaged polarization degrees from spectropolarimetric observations presented in \citet[][open symbols]{uno25}. Dotted lines show a linear fit to the NOT B- and R-band polarization degrees. An average systematic uncertainty from template-fitting procedure, as discussed in Section 2.1, is shown in the upper-left corner. \textit{Lower panel:} Multi-band polarization angle from our NOT observations (filled symbols) and band-averaged values from \citet[][open symbols]{uno25}.}
    \label{fig:polfig}
\end{figure}

We detect polarized optical emission from the source, with an increasing polarization degree and a varying polarization angle during the optical decay phase (Fig. \ref{fig:polfig}, Table \ref{tab:poltab}). The first observation, taken $\sim$6 days after the peak, shows a polarization degree of $\sim$1.5 per cent and polarization angle of $\sim$120$^{\circ}$ in the V- and R-bands, while in the B-band, the polarization degree is consistent with zero at the 1.7$\sigma$ level. 

In the second observation, taken at $\sim$20 days after the peak, the V-band polarization degree remains similar, while the degrees in the R and B bands increase to $\sim$2 and $\sim$3 per cent, respectively. This suggest a pivot in the polarization spectrum between the first two observing epochs---from red-dominated polarization (with V- and R-band polarization degrees $>$8$\sigma$ above the background, while the B-band is consistent with the background) to blue-dominated polarization (with the B-band polarization degree $\sim$4$\sigma$ higher than that in the V-band). At the same time, the polarization angle changes significantly, decreasing from $\sim$100-120$^{\circ}$ to $\sim$25-45$^{\circ}$. 

In the third observation, the B- and R-band polarization degrees are both around 3 per cent. The fourth observation shows the highest polarization levels, with $\sim$4 and $\sim$5 per cent in the B and R bands, respectively. The polarization angle remains similar to that in the second observing epoch. As noted in Section 2.2, the H$\alpha$ line flux can contribute more than 50 per cent of the total R-band flux, making it difficult to interpret the polarization spectrum during the latter two observing epochs. Over the four observations, the increase in polarization degree in the B and R bands is approximately linear (see dotted lines in Fig. \ref{fig:polfig}), with a growth rate of 0.11 per cent per day. 

However, $\sim$55 days after the peak, the polarization degree drops significantly. A 2$\sigma$ upper limit of 1.5 per cent in the V-band was obtained using the Multicolour OPTimised Optical Polarimeter (MOPTOP), mounted on the 2-m Liverpool Telescope \citep{charalampopoulos24}. Consistent results with this upper limit are reported by \citet{uno25}, whose data we have also included in Fig. \ref{fig:polfig} (open symbols). We convolved their q and u polarization spectra with the NOT filter transmission curves and calculated average polarization degrees and angles in the B, V, and R bands to enable direct comparison with our results. These show blue-dominated polarization with degrees below 2 per cent near the time of the MOPTOP observation. Their data obtained close in time to our first observing epoch also show comparable results, although our R- and V-band polarization degrees are higher by $\sim$1 per cent. This discrepancy may arise from differences in host subtraction and calibration methods.

\begin{table*}
    \centering
    \caption{Optical polarization parameters of AT 2023clx from our NOT observations. Both apparent and host-corrected values are shown.}    
    \begin{tabular}{ccccccccccc}
    \hline
      & & & & & & & \multicolumn{4}{|c|}{Host-corrected} \\
    \cline{8-11}
      MJD & MJD$-$59997.25 & Band & q & u & $\Pi$ (\%) & $\Theta$ ($^{\circ}$) & q & u & $\Pi$ (\%) & $\Theta$ ($^{\circ}$) \\
    \hline  
      60003.01 & 5.8 & B & $-0.27\pm0.11$ & $-0.06\pm0.11$ & $0.25\pm0.11$ & $96\pm13$ & $-0.23\pm0.12$ & $-0.05\pm0.12$ & $0.20\pm0.12$ & $99\pm19$ \\
      60017.19 & 19.9 & B & $-0.18\pm0.15$ & $0.99\pm0.15$ & $1.00\pm0.15$ & $50\pm4$ & $0.04\pm0.37$ & $3.32\pm0.36$ &  $3.30\pm0.36$ & $45\pm3$ \\
      60026.09 & 28.8 & B & $0.24\pm0.12$ & $1.24\pm0.12$ & $1.26\pm0.12$ & $40\pm3$ & $1.08\pm0.21$ & $2.86\pm0.29$ &   $3.05\pm0.25$ & $35\pm2$ \\
      60032.96 & 35.7 & B & $0.00\pm0.19$ & $0.95\pm0.19$ & $0.93\pm0.19$ & $45\pm6$ & $1.69\pm0.69$ & $3.34\pm0.51$ & $3.69\pm0.60$ & $32\pm5$ \\
    \hline  
      60003.01 & 5.8 & V & $-0.28\pm0.10$ & $-0.17\pm0.10$ & $0.32\pm0.10$ & $106\pm8$ & $-0.44\pm0.14$ & $-1.57\pm0.14$ & $1.63\pm0.14$ & $127\pm2$ \\
      60017.19 & 19.9 & V & $0.06\pm0.11$ & $0.40\pm0.11$ & $0.39\pm0.11$ & $41\pm7$ & $1.19\pm0.20$ & $1.28\pm0.20$ & $1.73\pm0.20$ & $24\pm3$ \\
    \hline
      60003.01 & 5.8 & R & $-0.18\pm0.07$ & $-0.04\pm0.07$ & $0.17\pm0.07$ & $96\pm11$ & $-0.63\pm0.10$ & $-1.18\pm0.10$ & $1.33\pm0.1$ & $121\pm2$ \\
      60017.19 & 19.9 & R & $0.01\pm0.07$ & $0.59\pm0.07$ & $0.59\pm0.07$ & $45\pm4$ & $0.32\pm0.16$ & $2.14\pm0.16$ &  $2.15\pm0.16$ & $41\pm2$ \\
      60026.09 & 28.8 & R & $0.19\pm0.07$ & $0.65\pm0.07$ & $0.67\pm0.07$ & $37\pm3$ & $1.60\pm0.33$ & $2.27\pm0.32$ & $2.76\pm0.33$ & $27\pm3$ \\
      60032.96 & 35.7 & R & $0.17\pm0.10$ & $0.59\pm0.10$ & $0.61\pm0.10$ & $37\pm5$ & $2.9\pm0.62$ & $3.84\pm0.68$ & $4.80\pm0.65$ & $26\pm4$ \\
    \hline  
    \end{tabular} 
    \label{tab:poltab}
\end{table*}

\section{Discussion}

The origin of optical polarization in TDEs remains an open question. A variable polarization degree and angle is difficult to reconcile with a simple reprocessing scenario unless the intrinsic emission geometry or scattering medium/surface evolves rapidly over time. However, tidal stream shocks are expected to change in both orientation and strength as the flow evolves \citep{shiokawa15,ryu23,steinberg24,price24}. This dynamic behavior can naturally lead to variability in the optical polarization parameters, particularly when the shocks are embedded in less dense material or when our line of sight is favorable for observing these shocks. 

In the following, we first check that other optical parameters of AT 2023clx conform with that expected from a shock scenario (Section 4.1). We then compare the evolution of the polarization parameters of AT 2023clx to those from AT 2020mot (Section 4.2) and note their scaled, temporal similarity. We discuss the potential evolution of the stellar stream shocks (Section 4.3), and its comparison to the observed from AT 2023clx. Finally, we discuss an alternative, reprocessing model that conforms with the observed polarization degree and its evolution (Section 4.4). 

\subsection{Shock scenario}

To estimate parameters for the shock scenario, we follow the formulation given in \citet{ryu20}, assuming $M_{\rm BH}=10^6\, M_{\odot}$ and $M_*=0.1\, M_{\odot}$ \citep{charalampopoulos24}. The apocenter distance is given as 

\begin{equation}
    a_0 = 6.5 \times 10^{14} M_{\rm BH,6}^{2/3} M_*^{2/9} \Xi^{-1} \, \mathrm{cm} \approx 6 \times 10^{14} \, \mathrm{cm}, 
\end{equation}

\noindent where $\Xi\approx0.64$ is a correction factor that accounts for stellar structure and relativistic effects, depending only on the stellar and black hole masses. The fallback time is: 

\begin{equation}
    t_0 = 3.2 \times 10^{6} M_{\rm BH,6}^{1/2} M_*^{1/3} \Xi^{-3/2} \, \mathrm{s} \approx 33.4 \, \mathrm{days},
\end{equation}

\noindent and the peak fallback rate is $\dot{M}_{\rm max} = M_*/(3t_0) = 0.36 M_{\odot}$/year. The maximum rate at which the outer shock dissipates energy is

\begin{equation}
    c_1 L_{\rm max} = 4.3 \times 10^{43} M_{\rm BH,6}^{-1/6} M_*^{4/9} \Xi^{5/2} \, \mathrm{erg} \, \mathrm{s}^{-1} \approx 0.5 \times 10^{43} \, \mathrm{erg} \, \mathrm{s}^{-1},
\end{equation}

\noindent where $c_1$ is an unknown scaling factor for the emission region. Associating all (or part) of the estimated bolometric luminosity at peak ($L_{\rm bol} = 1.6\times10^{43}$ erg s$^{-1}$; \citealt{charalampopoulos24}) with shock dissipation, we derive a lower limit for $c_1\geq0.31$. Consequently, the size of the emission region is $\sqrt{\Omega/4\pi}c_1a_0\geq 10^{14} \, \mathrm{cm}$, where $\Omega$ is the solid angle. We adopt $\Omega=2\pi$, similar to \citet{ryu20}, due to the emission surface likely being somewhat flattened. The corresponding peak blackbody temperature is: 

\begin{equation}
    T_{\rm max} = 2.3 \times 10^{4} c_1^{-3/4} M_{\rm BH,6}^{-3/8} \Xi^{9/8} \Big( \frac{\Omega}{2\pi} \Big)^{-1/4} \, \mathrm{K} \lesssim 24000 \, \mathrm{K}. 
\end{equation}

\noindent All these values are consistent with observations, which show a peak blackbody temperature of 21500$\pm$4000 K and a radius of $(4.6\pm0.6)\times10^{14}$ cm \citep{charalampopoulos24}. Thus, the observed black body luminosity and temperature of AT 2023clx are in line what would be expected from a shock scenario.

It is important to note that the fallback time does not necessarily correspond to the rise time of the flare. As discussed in \citet{charalampopoulos24}, the fast rise observed ($t_{\rm rise}=10.4\pm2.5$ days) could result from the disruption of a less centrally concentrated star, delaying the onset of emission. Interestingly, fast rises are also seen in recent three-dimensional radiation-hydrodynamic simulations, where shocks near pericenter drive the light curve evolution \citep{steinberg24}.   

\subsection{Comparison to AT 2020mot}

\begin{figure}
    \centering
    \includegraphics[width=1.0\linewidth]{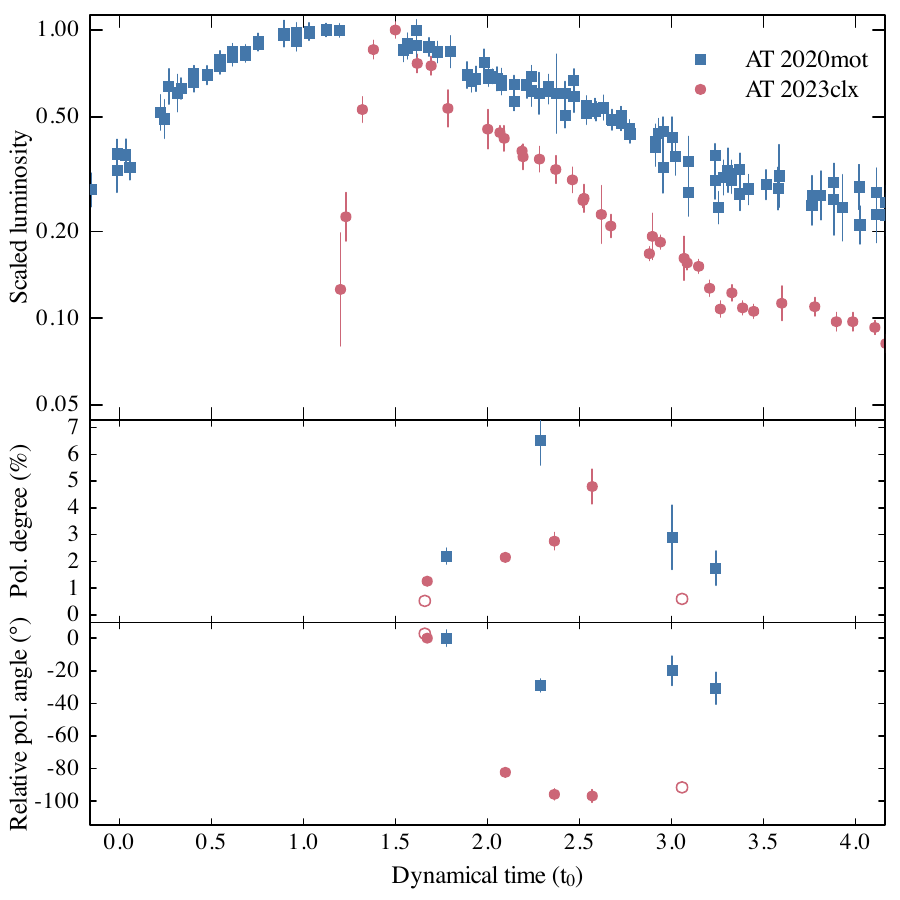}
    \caption{Comparison of the polarization properties of AT 2020mot (blue squares in all panels) and AT 2023clx (red circles in all panels) along their evolution of the fallback time (see Eq. 2). \textit{Upper panel:} Scaled (to peak) g-band luminosities. \textit{Middle panel:} R-band polarization degree from our NOT observations (filled symbols) and from \citet[][open symbols]{uno25}. Note that the polarization degree for AT 2020mot is not host-corrected for comparison purposes. The line depolarization effect for AT 2020mot is negligible \citep{liodakis23}. \textit{Lower panel:} R-band polarization angle from our NOT observations (filled symbols) and from \citet[][open symbols]{uno25}.}
    \label{fig:compare}
\end{figure}

The optical polarization properties of AT 2023clx closely resemble those observed in AT 2020mot \citep{liodakis23}. Fig. \ref{fig:compare} shows the evolution of the polarization properties for both sources over several fallback times (we aligned the peak emission to $1.5 \times t_0$; \citealt{ryu20} estimate the peak emission time as $1.5-2.0 \times t_0$)\footnote{Here, we rescaled the time axis as $t_{\rm dyn} = (t_{\rm MJD} - t_{\rm disruption})/t_0$, where $t_{\rm disruption} = t_{\rm peak, MJD} - 1.5 t_0$, and $t_0$ is calculated according to \citet{ryu20} for both systems separately.}. The polarization degree of AT 2023clx increases steadily (almost linearly) from the peak emission up to $\sim2.5 \times t_0$. At the same time, the polarization angle changes by $\sim$60--100$^{\circ}$. This behavior is identical to that observed in AT 2020mot, where the polarization degree began rising after $1.5 \times t_0$, peaking around $\sim2.5 \times t_0$. During this period, the polarization angle changed by $\sim 30^{\circ}$. After reaching its peak polarization degree, the polarization of AT 2020mot decreased beyond $3.0 \times t_0$, while the polarization angle remained constant. For AT 2023clx, although we do not have NOT observations during the expected decline, there is a clear drop of the polarization degree after $3.0 \times t_0$ based on the spectropolarimetry presented in \citet{uno25} and the MOPTOP upper limit.

The only notable difference between the optical polarization properties of AT 2020mot and AT 2023clx is the magnitude of the polarization degree: AT 2020mot displayed a host-corrected peak of $\sim$25 per cent, while AT 2023clx exhibited a maximum of $\sim$5 per cent. This discrepancy could be attributed to line-of-sight scattering, as AT 2023clx shows more spectral lines in its optical spectra than AT 2020mot, suggesting the presence of additional material that dilutes the observed polarized emission. It is possible that, in the case of AT 2020mot, we had an unusually clear view of the tidal stream shocks. Alternatively, the ordering of the magnetic field of the shocks could be more chaotic in AT 2023clx than AT 2020mot effectively reducing the observed polarization degree.  

We note that outside AT 2020mot multiple epoch polarimetry with more than two observing epochs has not been performed for other TDEs with the exception of AT 2022fpx \citep{koljonen24}. However, this system is an extreme coronal line emitter, with very slow outburst decay, and therefore distinct from \textit{bona fide} TDEs. Therefore, this prevents us making population-wide conclusions on the typical optical polarimetric evolution of TDEs.

\subsection{Shock evolution}

\citet{liodakis23} explained the evolution of shocks based on the calculations presented in \citet{shiokawa15} for the tidal disruption of a white dwarf by an intermediate-mass black hole, with extensions for higher black hole masses as discussed in \citet{piran15}. According to these studies, the dominant mechanism of shock dissipation evolves over time, beginning with the nozzle shock and later transitioning to outer shocks (reverse and forward shocks) at approximately $2-3 \times t_0$. The three-dimensional radiation-hydrodynamic simulation by \citet{steinberg24} largely agree with this scenario during the early stages, although they propose rapid circularization around the time of peak emission and highlight the significant role of stream-disk shocks rather than stream-stream collisions. 

The optical lightcurve of AT 2023clx also exhibits notable changes that align with the approximate shock phases described above. Between the peak and $2 \times t_0$, the source underwent a `rapid cooling' phase \citep{charalampopoulos24}, characterized by a steep decrease in blackbody temperature and a corresponding increase in blackbody radius. This behavior could indicate the dominance of outer shocks occurring at larger radii within the stellar debris. Following this phase, the lightcurve slope became more shallow, coinciding with an increase in the polarization degree and a shift in the polarization angle. This shallower slope may result from enhanced dissipation caused by either outer shocks or stream-disk shocks.

After $3 \times t_0$, the lightcurve flattened further, potentially indicating the emergence of an additional source of optical emission, possibly associated with the onset of accretion disk emission. Indeed, X-ray emission from AT 2023clx was detected $\sim$100 days post-peak (or $4.4 \times t_0$), signaling the formation of an accretion disk \citep{hoogendam24}. In the shock model, the formation of the accretion disk is expected to occur between 3 and $10 \times t_0$ \citep{shiokawa15}.

\subsection{Collisionally-induced outflow scenario}

\citet{charalampopoulos23} proposed a model for optical polarization in TDEs based on the reprocessing scenario of \citet{lu20}. In this model, unbound shocked gas originating from the intersection point of debris streams, offset from the black hole, reprocesses the X-ray emission from the accretion flow into ultraviolet and optical wavelengths. The polarization arises from electron scattering within the stellar debris that typically produces wavelength-independent polarization. However, we note that in at least some observing epochs, the measured polarization is wavelength-dependent (see Section 3; see also \citealt{uno25}).

The parameters for AT 2023clx align well with those of case B in their model, which assumes a disrupted star mass $M_*=0.1M_{\odot}$ and a fallback rate of $\dot{M}=0.3M_{\odot}$/year. Simulations for this case predict a polarization degree as high as 14 per cent, comfortably encompassing our observed maximum of $\sim$5 per cent. High polarization degrees in the model are generally associated with large distances between the black hole and the stream intersection point, small opening angles of the intrinsic emission, and intermediate viewing angles of the system (60$^{\circ}$--90$^{\circ}$). The model also predicts a rising polarization degree as the density and optical depth of the expanding debris decrease, followed by a decline in polarization when the optical depth becomes too low to sustain significant electron scattering (expected at $t\gtrsim3\times t_0$ in our case). 

Thus, the collisionally-induced outflow model also successfully explains both the evolution and magnitude of the polarization degree in AT 2023clx. However, currently it does not predict polarization angles or their evolution throughout the TDE, preventing direct comparison with observations in this aspect. The above challenge is also shared by other reprocessing-based scenarios.
 
\section{Conclusions}

Our observations of AT 2023clx revealed a variable polarization degree and angle, highlighting the dynamic evolution of the emission geometry and scattering media during the optical peak and subsequent decay. The evolution of polarization properties of AT 2023clx closely resemble those observed in AT 2020mot, supporting the hypothesis that tidal stream shocks play a significant role in shaping the optical emission in these events.

Future work should focus on resolving the discrepancies between observed and predicted polarization properties, particularly the polarization angle variability. High-cadence, multi-wavelength polarization observations, combined with improved theoretical modeling, will be crucial to untangle the contributions of shocks, reprocessing, and accretion flows in shaping the emission from TDEs.

\section*{Acknowledgements}

We sincerely thank Panagiotis Charalampopoulos for providing the NOT V-band polarization data (obtained under programme P66-019) and the host-subtracted spectra taken close to our NOT observations. We also thank the anonymous referee for comments that improved the manuscript.

This project has received funding from the European Research Council (ERC) under the European Union’s Horizon 2020 research and innovation programme (grant agreement No. 101002352, PI: M. Linares).

I.L was funded by the European Union ERC-2022-STG - BOOTES - 101076343. Views and opinions expressed are however those of the author(s) only and do not necessarily reflect those of the European Union or the European Research Council Executive Agency. Neither the European Union nor the granting authority can be held responsible for them.

The data presented here were obtained with ALFOSC, which is provided by the Instituto de Astrofisica de Andalucia (IAA) under a joint agreement with the University of Copenhagen and NOT.

Based on observations made with the Nordic Optical Telescope, owned in collaboration by the University of Turku and Aarhus University, and operated jointly by Aarhus University, the University of Turku and the University of Oslo, representing Denmark, Finland and Norway, the University of Iceland and Stockholm University at the Observatorio del Roque de los Muchachos, La Palma, Spain, of the Instituto de Astrofisica de Canarias.

\section*{Data Availability}

All data is freely accessible in the NOT archive (https://www.not.iac.es/archive/).



\bibliographystyle{mnras}
\bibliography{bibliography} 

\begin{thebibliography}{}
\makeatletter
\relax
\def\mn@urlcharsother{\let\do\@makeother \do\$\do\&\do\#\do\^\do\_\do\%\do\~}
\def\mn@doi{\begingroup\mn@urlcharsother \@ifnextchar [ {\mn@doi@}
  {\mn@doi@[]}}
\def\mn@doi@[#1]#2{\def\@tempa{#1}\ifx\@tempa\@empty \href
  {http://dx.doi.org/#2} {doi:#2}\else \href {http://dx.doi.org/#2} {#1}\fi
  \endgroup}
\def\mn@eprint#1#2{\mn@eprint@#1:#2::\@nil}
\def\mn@eprint@arXiv#1{\href {http://arxiv.org/abs/#1} {{\tt arXiv:#1}}}
\def\mn@eprint@dblp#1{\href {http://dblp.uni-trier.de/rec/bibtex/#1.xml}
  {dblp:#1}}
\def\mn@eprint@#1:#2:#3:#4\@nil{\def\@tempa {#1}\def\@tempb {#2}\def\@tempc
  {#3}\ifx \@tempc \@empty \let \@tempc \@tempb \let \@tempb \@tempa \fi \ifx
  \@tempb \@empty \def\@tempb {arXiv}\fi \@ifundefined
  {mn@eprint@\@tempb}{\@tempb:\@tempc}{\expandafter \expandafter \csname
  mn@eprint@\@tempb\endcsname \expandafter{\@tempc}}}

\bibitem[\protect\citeauthoryear{{Albareti} et~al.,}{{Albareti}
  et~al.}{2017}]{albareti17}
{Albareti} F.~D.,  et~al., 2017, \mn@doi [\apjs] {10.3847/1538-4365/aa8992},
  \href {https://ui.adsabs.harvard.edu/abs/2017ApJS..233...25A} {233, 25}

\bibitem[\protect\citeauthoryear{{Angel} \& {Stockman}}{{Angel} \&
  {Stockman}}{1980}]{angel80}
{Angel} J.~R.~P.,  {Stockman} H.~S.,  1980, \mn@doi [\araa]
  {10.1146/annurev.aa.18.090180.001541}, \href
  {https://ui.adsabs.harvard.edu/abs/1980ARA&A..18..321A} {18, 321}

\bibitem[\protect\citeauthoryear{{Cendes} et~al.,}{{Cendes}
  et~al.}{2024}]{cendes24}
{Cendes} Y.,  et~al., 2024, \mn@doi [\apj] {10.3847/1538-4357/ad5541}, \href
  {https://ui.adsabs.harvard.edu/abs/2024ApJ...971..185C} {971, 185}

\bibitem[\protect\citeauthoryear{{Charalampopoulos}, {Bulla}, {Bonnerot}  \&
  {Leloudas}}{{Charalampopoulos} et~al.}{2023}]{charalampopoulos23}
{Charalampopoulos} P.,  {Bulla} M.,  {Bonnerot} C.,   {Leloudas} G.,  2023,
  \mn@doi [\aap] {10.1051/0004-6361/202245014}, \href
  {https://ui.adsabs.harvard.edu/abs/2023A&A...670A.150C} {670, A150}

\bibitem[\protect\citeauthoryear{{Charalampopoulos} et~al.,}{{Charalampopoulos}
  et~al.}{2024}]{charalampopoulos24}
{Charalampopoulos} P.,  et~al., 2024, \mn@doi [\aap]
  {10.1051/0004-6361/202449296}, \href
  {https://ui.adsabs.harvard.edu/abs/2024A&A...689A.350C} {689, A350}

\bibitem[\protect\citeauthoryear{{Evans} \& {Kochanek}}{{Evans} \&
  {Kochanek}}{1989}]{evans89}
{Evans} C.~R.,  {Kochanek} C.~S.,  1989, \mn@doi [\apjl] {10.1086/185567},
  \href {https://ui.adsabs.harvard.edu/abs/1989ApJ...346L..13E} {346, L13}

\bibitem[\protect\citeauthoryear{{Gezari} et~al.,}{{Gezari}
  et~al.}{2012}]{gezari12}
{Gezari} S.,  et~al., 2012, \mn@doi [\nat] {10.1038/nature10990}, \href
  {https://ui.adsabs.harvard.edu/abs/2012Natur.485..217G} {485, 217}

\bibitem[\protect\citeauthoryear{{Gezari}, {Cenko}  \& {Arcavi}}{{Gezari}
  et~al.}{2017}]{gezari17}
{Gezari} S.,  {Cenko} S.~B.,   {Arcavi} I.,  2017, \mn@doi [\apjl]
  {10.3847/2041-8213/aaa0c2}, \href
  {https://ui.adsabs.harvard.edu/abs/2017ApJ...851L..47G} {851, L47}

\bibitem[\protect\citeauthoryear{{Guillochon}, {Manukian}  \&
  {Ramirez-Ruiz}}{{Guillochon} et~al.}{2014}]{guillochon14}
{Guillochon} J.,  {Manukian} H.,   {Ramirez-Ruiz} E.,  2014, \mn@doi [\apj]
  {10.1088/0004-637X/783/1/23}, \href
  {https://ui.adsabs.harvard.edu/abs/2014ApJ...783...23G} {783, 23}

\bibitem[\protect\citeauthoryear{{Guolo}, {Gezari}, {Yao}, {van Velzen},
  {Hammerstein}, {Cenko}  \& {Tokayer}}{{Guolo} et~al.}{2024}]{guolo24}
{Guolo} M.,  {Gezari} S.,  {Yao} Y.,  {van Velzen} S.,  {Hammerstein} E.,
  {Cenko} S.~B.,   {Tokayer} Y.~M.,  2024, \mn@doi [\apj]
  {10.3847/1538-4357/ad2f9f}, \href
  {https://ui.adsabs.harvard.edu/abs/2024ApJ...966..160G} {966, 160}

\bibitem[\protect\citeauthoryear{{Ho}}{{Ho}}{2008}]{ho08}
{Ho} L.~C.,  2008, \mn@doi [\araa] {10.1146/annurev.astro.45.051806.110546},
  \href {https://ui.adsabs.harvard.edu/abs/2008ARA&A..46..475H} {46, 475}

\bibitem[\protect\citeauthoryear{{Hoogendam} et~al.,}{{Hoogendam}
  et~al.}{2024}]{hoogendam24}
{Hoogendam} W.~B.,  et~al., 2024, \mn@doi [\mnras] {10.1093/mnras/stae1121},
  \href {https://ui.adsabs.harvard.edu/abs/2024MNRAS.530.4501H} {530, 4501}

\bibitem[\protect\citeauthoryear{{Horesh}, {Cenko}  \& {Arcavi}}{{Horesh}
  et~al.}{2021}]{horesh21}
{Horesh} A.,  {Cenko} S.~B.,   {Arcavi} I.,  2021, \mn@doi [Nature Astronomy]
  {10.1038/s41550-021-01300-8}, \href
  {https://ui.adsabs.harvard.edu/abs/2021NatAs...5..491H} {5, 491}

\bibitem[\protect\citeauthoryear{{Hovatta} \& {Lindfors}}{{Hovatta} \&
  {Lindfors}}{2019}]{hovatta19}
{Hovatta} T.,  {Lindfors} E.,  2019, \mn@doi [\nar]
  {10.1016/j.newar.2020.101541}, \href
  {https://ui.adsabs.harvard.edu/abs/2019NewAR..8701541H} {87, 101541}

\bibitem[\protect\citeauthoryear{{Jiang}, {Guillochon}  \& {Loeb}}{{Jiang}
  et~al.}{2016}]{jiang16}
{Jiang} Y.-F.,  {Guillochon} J.,   {Loeb} A.,  2016, \mn@doi [\apj]
  {10.3847/0004-637X/830/2/125}, \href
  {https://ui.adsabs.harvard.edu/abs/2016ApJ...830..125J} {830, 125}

\bibitem[\protect\citeauthoryear{{Johansson}, {Meynardie}, {Chu}  \&
  {Fremling}}{{Johansson} et~al.}{2023}]{johansson23}
{Johansson} J.,  {Meynardie} W.,  {Chu} M.,   {Fremling} C.,  2023, Transient
  Name Server Classification Report, \href
  {https://ui.adsabs.harvard.edu/abs/2023TNSCR1075....1J} {2023-1075, 1}

\bibitem[\protect\citeauthoryear{{Kewley}, {Zahid}, {Geller}, {Dopita}, {Hwang}
   \& {Fabricant}}{{Kewley} et~al.}{2015}]{kewley15}
{Kewley} L.~J.,  {Zahid} H.~J.,  {Geller} M.~J.,  {Dopita} M.~A.,  {Hwang}
  H.~S.,   {Fabricant} D.,  2015, \mn@doi [\apjl]
  {10.1088/2041-8205/812/2/L20}, \href
  {https://ui.adsabs.harvard.edu/abs/2015ApJ...812L..20K} {812, L20}

\bibitem[\protect\citeauthoryear{{Koljonen} et~al.,}{{Koljonen}
  et~al.}{2024}]{koljonen24}
{Koljonen} K. I.~I.,  et~al., 2024, \mn@doi [\mnras] {10.1093/mnras/stae1466},
  \href {https://ui.adsabs.harvard.edu/abs/2024MNRAS.532..112K} {532, 112}

\bibitem[\protect\citeauthoryear{{Landi Degl'Innocenti}, {Bagnulo}  \&
  {Fossati}}{{Landi Degl'Innocenti} et~al.}{2007}]{landi07}
{Landi Degl'Innocenti} E.,  {Bagnulo} S.,   {Fossati} L.,  2007, in {Sterken}
  C.,  ed.,  Astronomical Society of the Pacific Conference Series Vol. 364,
  The Future of Photometric, Spectrophotometric and Polarimetric
  Standardization. p.~495 (\mn@eprint {arXiv} {astro-ph/0610262})

\bibitem[\protect\citeauthoryear{{Leloudas} et~al.,}{{Leloudas}
  et~al.}{2022}]{leloudas22}
{Leloudas} G.,  et~al., 2022, \mn@doi [Nature Astronomy]
  {10.1038/s41550-022-01767-z}, \href
  {https://ui.adsabs.harvard.edu/abs/2022NatAs...6.1193L} {6, 1193}

\bibitem[\protect\citeauthoryear{{Liodakis} et~al.,}{{Liodakis}
  et~al.}{2023}]{liodakis23}
{Liodakis} I.,  et~al., 2023, \mn@doi [Science] {10.1126/science.abj9570},
  \href {https://ui.adsabs.harvard.edu/abs/2023Sci...380..656L} {380, 656}

\bibitem[\protect\citeauthoryear{{Lodato} \& {Rossi}}{{Lodato} \&
  {Rossi}}{2011}]{lodato11}
{Lodato} G.,  {Rossi} E.~M.,  2011, \mn@doi [\mnras]
  {10.1111/j.1365-2966.2010.17448.x}, \href
  {https://ui.adsabs.harvard.edu/abs/2011MNRAS.410..359L} {410, 359}

\bibitem[\protect\citeauthoryear{{Loeb} \& {Ulmer}}{{Loeb} \&
  {Ulmer}}{1997}]{loeb97}
{Loeb} A.,  {Ulmer} A.,  1997, \mn@doi [\apj] {10.1086/304814}, \href
  {https://ui.adsabs.harvard.edu/abs/1997ApJ...489..573L} {489, 573}

\bibitem[\protect\citeauthoryear{{Lu} \& {Bonnerot}}{{Lu} \&
  {Bonnerot}}{2020}]{lu20}
{Lu} W.,  {Bonnerot} C.,  2020, \mn@doi [\mnras] {10.1093/mnras/stz3405}, \href
  {https://ui.adsabs.harvard.edu/abs/2020MNRAS.492..686L} {492, 686}

\bibitem[\protect\citeauthoryear{{Metzger} \& {Stone}}{{Metzger} \&
  {Stone}}{2016}]{metzger16}
{Metzger} B.~D.,  {Stone} N.~C.,  2016, \mn@doi [\mnras]
  {10.1093/mnras/stw1394}, \href
  {https://ui.adsabs.harvard.edu/abs/2016MNRAS.461..948M} {461, 948}

\bibitem[\protect\citeauthoryear{{Mummery}, {van Velzen}, {Nathan}, {Ingram},
  {Hammerstein}, {Fraser-Taliente}  \& {Balbus}}{{Mummery}
  et~al.}{2024}]{mummery24}
{Mummery} A.,  {van Velzen} S.,  {Nathan} E.,  {Ingram} A.,  {Hammerstein} E.,
  {Fraser-Taliente} L.,   {Balbus} S.,  2024, \mn@doi [\mnras]
  {10.1093/mnras/stad3001}, \href
  {https://ui.adsabs.harvard.edu/abs/2024MNRAS.527.2452M} {527, 2452}

\bibitem[\protect\citeauthoryear{{Mundell} et~al.,}{{Mundell}
  et~al.}{2013}]{mundell13}
{Mundell} C.~G.,  et~al., 2013, \mn@doi [\nat] {10.1038/nature12814}, \href
  {https://ui.adsabs.harvard.edu/abs/2013Natur.504..119M} {504, 119}

\bibitem[\protect\citeauthoryear{{Parkinson}, {Knigge}, {Matthews}, {Long},
  {Higginbottom}, {Sim}  \& {Mangham}}{{Parkinson} et~al.}{2022}]{parkinson22}
{Parkinson} E.~J.,  {Knigge} C.,  {Matthews} J.~H.,  {Long} K.~S.,
  {Higginbottom} N.,  {Sim} S.~A.,   {Mangham} S.~W.,  2022, \mn@doi [\mnras]
  {10.1093/mnras/stac027}, \href
  {https://ui.adsabs.harvard.edu/abs/2022MNRAS.510.5426P} {510, 5426}

\bibitem[\protect\citeauthoryear{{Pasham}, {Cenko}, {Sadowski}, {Guillochon},
  {Stone}, {van Velzen}  \& {Cannizzo}}{{Pasham} et~al.}{2017}]{pasham17}
{Pasham} D.~R.,  {Cenko} S.~B.,  {Sadowski} A.,  {Guillochon} J.,  {Stone}
  N.~C.,  {van Velzen} S.,   {Cannizzo} J.~K.,  2017, \mn@doi [\apjl]
  {10.3847/2041-8213/aa6003}, \href
  {https://ui.adsabs.harvard.edu/abs/2017ApJ...837L..30P} {837, L30}

\bibitem[\protect\citeauthoryear{{Patra}, {Lu}, {Brink}, {Yang}, {Filippenko}
  \& {Vasylyev}}{{Patra} et~al.}{2022}]{patra22}
{Patra} K.~C.,  {Lu} W.,  {Brink} T.~G.,  {Yang} Y.,  {Filippenko} A.~V.,
  {Vasylyev} S.~S.,  2022, \mn@doi [\mnras] {10.1093/mnras/stac1727}, \href
  {https://ui.adsabs.harvard.edu/abs/2022MNRAS.515..138P} {515, 138}

\bibitem[\protect\citeauthoryear{{Piran}, {Svirski}, {Krolik}, {Cheng}  \&
  {Shiokawa}}{{Piran} et~al.}{2015}]{piran15}
{Piran} T.,  {Svirski} G.,  {Krolik} J.,  {Cheng} R.~M.,   {Shiokawa} H.,
  2015, \mn@doi [\apj] {10.1088/0004-637X/806/2/164}, \href
  {https://ui.adsabs.harvard.edu/abs/2015ApJ...806..164P} {806, 164}

\bibitem[\protect\citeauthoryear{{Price}, {Liptai}, {Mandel}, {Shepherd},
  {Lodato}  \& {Levin}}{{Price} et~al.}{2024}]{price24}
{Price} D.~J.,  {Liptai} D.,  {Mandel} I.,  {Shepherd} J.,  {Lodato} G.,
  {Levin} Y.,  2024, \mn@doi [\apjl] {10.3847/2041-8213/ad6862}, \href
  {https://ui.adsabs.harvard.edu/abs/2024ApJ...971L..46P} {971, L46}

\bibitem[\protect\citeauthoryear{{Ramos Padilla}, {Ashby}, {Smith},
  {Mart{\'\i}nez-Galarza}, {Beverage}, {Dietrich}, {Higuera-G.}  \&
  {Weiner}}{{Ramos Padilla} et~al.}{2020}]{ramospadilla20}
{Ramos Padilla} A.~F.,  {Ashby} M.~L.~N.,  {Smith} H.~A.,
  {Mart{\'\i}nez-Galarza} J.~R.,  {Beverage} A.~G.,  {Dietrich} J.,
  {Higuera-G.} M.-A.,   {Weiner} A.~S.,  2020, \mn@doi [\mnras]
  {10.1093/mnras/staa2813}, \href
  {https://ui.adsabs.harvard.edu/abs/2020MNRAS.499.4325R} {499, 4325}

\bibitem[\protect\citeauthoryear{{Rees}}{{Rees}}{1988}]{rees88}
{Rees} M.~J.,  1988, \mn@doi [\nat] {10.1038/333523a0}, \href
  {https://ui.adsabs.harvard.edu/abs/1988Natur.333..523R} {333, 523}

\bibitem[\protect\citeauthoryear{{Roth}, {Kasen}, {Guillochon}  \&
  {Ramirez-Ruiz}}{{Roth} et~al.}{2016}]{roth16}
{Roth} N.,  {Kasen} D.,  {Guillochon} J.,   {Ramirez-Ruiz} E.,  2016, \mn@doi
  [\apj] {10.3847/0004-637X/827/1/3}, \href
  {https://ui.adsabs.harvard.edu/abs/2016ApJ...827....3R} {827, 3}

\bibitem[\protect\citeauthoryear{{Ryu}, {Krolik}  \& {Piran}}{{Ryu}
  et~al.}{2020}]{ryu20}
{Ryu} T.,  {Krolik} J.,   {Piran} T.,  2020, \mn@doi [\apj]
  {10.3847/1538-4357/abbf4d}, \href
  {https://ui.adsabs.harvard.edu/abs/2020ApJ...904...73R} {904, 73}

\bibitem[\protect\citeauthoryear{{Ryu}, {Krolik}, {Piran}, {Noble}  \&
  {Avara}}{{Ryu} et~al.}{2023}]{ryu23}
{Ryu} T.,  {Krolik} J.,  {Piran} T.,  {Noble} S.~C.,   {Avara} M.,  2023,
  \mn@doi [\apj] {10.3847/1538-4357/acf5de}, \href
  {https://ui.adsabs.harvard.edu/abs/2023ApJ...957...12R} {957, 12}

\bibitem[\protect\citeauthoryear{{Sfaradi}, {Horesh}, {Bright}, {Fender},
  {Rhodes}, {Green}  \& {Titterington}}{{Sfaradi} et~al.}{2023}]{sfaradi23}
{Sfaradi} I.,  {Horesh} A.,  {Bright} J.,  {Fender} R.,  {Rhodes} L.,  {Green}
  D.,   {Titterington} D.,  2023, The Astronomer's Telegram, \href
  {https://ui.adsabs.harvard.edu/abs/2023ATel15918....1S} {15918, 1}

\bibitem[\protect\citeauthoryear{{Shiokawa}, {Krolik}, {Cheng}, {Piran}  \&
  {Noble}}{{Shiokawa} et~al.}{2015}]{shiokawa15}
{Shiokawa} H.,  {Krolik} J.~H.,  {Cheng} R.~M.,  {Piran} T.,   {Noble} S.~C.,
  2015, \mn@doi [\apj] {10.1088/0004-637X/804/2/85}, \href
  {https://ui.adsabs.harvard.edu/abs/2015ApJ...804...85S} {804, 85}

\bibitem[\protect\citeauthoryear{{Simmons} \& {Stewart}}{{Simmons} \&
  {Stewart}}{1985}]{simmons85}
{Simmons} J.~F.~L.,  {Stewart} B.~G.,  1985, \aap, \href
  {https://ui.adsabs.harvard.edu/abs/1985A&A...142..100S} {142, 100}

\bibitem[\protect\citeauthoryear{{Steinberg} \& {Stone}}{{Steinberg} \&
  {Stone}}{2024}]{steinberg24}
{Steinberg} E.,  {Stone} N.~C.,  2024, \mn@doi [\nat]
  {10.1038/s41586-023-06875-y}, \href
  {https://ui.adsabs.harvard.edu/abs/2024Natur.625..463S} {625, 463}

\bibitem[\protect\citeauthoryear{{Strubbe} \& {Quataert}}{{Strubbe} \&
  {Quataert}}{2009}]{strubbe09}
{Strubbe} L.~E.,  {Quataert} E.,  2009, \mn@doi [\mnras]
  {10.1111/j.1365-2966.2009.15599.x}, \href
  {https://ui.adsabs.harvard.edu/abs/2009MNRAS.400.2070S} {400, 2070}

\bibitem[\protect\citeauthoryear{{Taguchi}, {Uno}, {Nagao}  \&
  {Maeda}}{{Taguchi} et~al.}{2023}]{taguchi23}
{Taguchi} K.,  {Uno} K.,  {Nagao} T.,   {Maeda} K.,  2023, Transient Name
  Server Classification Report, \href
  {https://ui.adsabs.harvard.edu/abs/2023TNSCR.438....1T} {2023-438, 1}

\bibitem[\protect\citeauthoryear{{Uno} et~al.,}{{Uno} et~al.}{2025}]{uno25}
{Uno} K.,  et~al., 2025, \mn@doi [arXiv e-prints] {10.48550/arXiv.2503.19024},
  \href {https://ui.adsabs.harvard.edu/abs/2025arXiv250319024U} {p.
  arXiv:2503.19024}

\bibitem[\protect\citeauthoryear{{Yan} \& {Blanton}}{{Yan} \&
  {Blanton}}{2012}]{yan12}
{Yan} R.,  {Blanton} M.~R.,  2012, \mn@doi [\apj] {10.1088/0004-637X/747/1/61},
  \href {https://ui.adsabs.harvard.edu/abs/2012ApJ...747...61Y} {747, 61}

\bibitem[\protect\citeauthoryear{{Zhu}, {Jiang}, {Wang}, {Huang}, {Lin}, {Wang}
   \& {Wang}}{{Zhu} et~al.}{2023}]{zhu23}
{Zhu} J.,  {Jiang} N.,  {Wang} T.,  {Huang} S.,  {Lin} Z.,  {Wang} Y.,   {Wang}
  J.-G.,  2023, \mn@doi [\apjl] {10.3847/2041-8213/ace625}, \href
  {https://ui.adsabs.harvard.edu/abs/2023ApJ...952L..35Z} {952, L35}

\bibitem[\protect\citeauthoryear{{van Velzen} et~al.,}{{van Velzen}
  et~al.}{2021}]{vanvelzen21}
{van Velzen} S.,  et~al., 2021, \mn@doi [\apj] {10.3847/1538-4357/abc258},
  \href {https://ui.adsabs.harvard.edu/abs/2021ApJ...908....4V} {908, 4}

\makeatother
\end{thebibliography}




%
%


\bsp	
\label{lastpage}
\end{document}